\documentclass[aps, 12pt, amsmath, amssymb, noshowpacs, preprint,preprintnumbers]{revtex4-1}
\usepackage{amsmath}
\usepackage{amsfonts}
\usepackage{amssymb}
\usepackage{textcomp}
\usepackage{graphicx}
\usepackage{bm}
\usepackage{color}

\usepackage{epsfig}
\usepackage{subfigure}
\usepackage{amsfonts,amsmath,amssymb}

\newcommand{\be}{\begin{equation}}
\newcommand{\ee}{\end{equation}}
\newcommand{\ba}{\begin{eqnarray}}
\newcommand{\ea}{\end{eqnarray}}
\newcommand{\ben}{\begin{enumerate}}
\newcommand{\een}{\end{enumerate}}

\newcommand{\lb}{\left(}
\newcommand{\rb}{\right)}

\newcommand{\ld}{\left.}

\newcommand{\rv}{\right|}
\newcommand{\lbr}{\left[}
\newcommand{\rbr}{\right]}

\newcommand{\p}{\partial}

\newcommand{\la}{\langle}
\newcommand{\ra}{\rangle}

\newcommand{\rar}{\rightarrow}

\begin{document}
\title{Wilson loops in holographic models with a gluon condensate}
\author{P.~N.~Kopnin, A.~Krikun}
\affiliation{MIPT, Moscow, Russia}
\affiliation{ITEP, Moscow, Russia}
\preprint{ITEP-TH-19/11}
\begin{abstract}
The aim of this work is to study the holographic dual to the gauge theory with a nonzero gluon condensate. We check for consistency the holographic way of describing the condensate and calculate the expectation value of a small Wilson loop in the presence of the gluon condensate, thus obtaining the relevant coefficient in the operator product expansion of the small loop in different holographic models. We also study the effect of the condensate on the Gross-Ooguri phase transition in the correlator of two circular Wilson loops in parallel and concentric configurations. In the numerical study of the concentric case, we find that the phase transition changes its order when the size of the loops is of order of the gluon condensate. We report this change of the phase transition order to be a new effect in Wilson loop correlators.
\end{abstract}

\maketitle

\section{Introduction}
Holographic models provide a powerful machinery for the nonperturbative description of gauge theories in the strong coupling regime. Formulated first as a method of description of the strongly coupled supersymmetric conformal gauge theory N=4 SYM \cite{Maldacena_D3,Gubser-Klebanov,Witten_AdS}, the AdS/CFT correspondence was then used in a phenomenological approach to study the real QCD \cite{Erlich}. Its deformations breaking the supersymmetry and conformal symmetry were proposed  as well in an effort to consistently reduce the AdS/CFT correspondence to the description of the a QCD-like theory \cite{Witten_thermal,Polchinski-Strassler,Gubser}. 

One of the main manifestations of the broken conformal symmetry in QCD is the nonzero gluon condensate, the nonperturbative vacuum expectation value of the scalar gluon operator
\begin{equation}
\la \alpha_s {\rm tr}(G^2) \ra,
\end{equation}
introduced in the framework of QCD sum rules in \cite{SVZ}. This value plays a major role in the QCD sum  rules method and has a variety of phenomenological applications. For instance, one can show, that the $\rho$ and other meson masses are proportional to $\la \alpha_s {\rm tr}(G^2) \ra^{1/4}$, and the energy of QCD vacuum equals $\epsilon = -\frac{b}{32} \la \frac{\alpha_s}{\pi} {\rm tr}(G^2) \ra$. Therefore, there are strong reasons to believe that in the real QCD the vacuum expectation value under consideration does not vanish. The phenomenological value of the gluon condensate which is usually used is of order of
\begin{equation}
\label{trGG}
\la \alpha_s {\rm tr}(G^2) \ra \sim (200\ {\rm MeV})^4.
\end{equation}
In the framework of the QCD sum rules the gluon condensate is a free parameter, which can be adjusted in order to get better phenomenological results. Later, in the 1980-ies, in order to determine whether the condensate really exists, it was measured in the fundamental theory via lattice calculations. For this purpose one can study the expectation value of a small Wilson loop operator
\begin{equation}
\label{loop}
W(\mathcal{C}) = \frac{1}{N_c} \la tr \ P\exp[\oint_\mathcal{C} i g A_\mu dx^\mu ] \ra .
\end{equation}
As was shown in \cite{Shifman}, a sufficiently small Wilson loop can be presented in the form of an operator product expansion, which looks like (omitting the input of the perturbation theory)
\begin{equation}
\label{shif}
W(\mathcal{C}) = 1 - \frac{1}{48} \frac{\la \alpha_s {\rm tr}(G^2) \ra}{N_c} S^2 + \dots,
\end{equation}
where $S$ is the area of the Wilson contour $\mathcal{C}$. This relation was a basis for the lattice study of the gluon condensate, which has led to a reasonable value, while the procedure of subtracting the perturbative contribution into the Wilson loop is not well defined \cite{Banks,Di Giacomo}.

In the gauge/gravity duality the gluon condensate can be calculated in two at first sight unrelated ways. First of all, as the gluon condensate is a vacuum expectation value of a certain operator, it can be extracted as a coefficient in front of the normalizable mode of its dual field in the supergravity \cite{Gubser-Klebanov}. In fact there is a general way to show that the field dual to the ${\rm tr}(G^2)$ operator is the dilaton \cite{Ooguri-Oz}. That means that to describe a field theory with a nonzero gluon condensate we need to consider the supergravity background with a nontrivial normalizable mode of the dilaton field
\begin{equation}
\label{nontriv_dil}
\phi = \phi_0 + \phi_4 z^4, 
\end{equation}
where $z$ is a distance to the boundary of the holographic background.

On the other hand, one can study the area of the minimal surface, spanned on the circular contour of a small radius on the boundary, which is related to the expectation value of the field-theoretic Wilson loop on this contour, and get the value of the gluon condensate in the spirit of lattice calculations, as it was done in \cite{Andreev:2007vn}. In this approach the gluon condensate is extracted from the OPE of the Wilson loop operator. In fact, the relation (\ref{shif}) is valid only in the small coupling limit, while the holographic calculation implies a strong coupling. That means that we need the version of the Wilson loop OPE at a strong coupling, which was obtained in \cite{Berenstein-Maldacena} in the AdS/CFT approach. Most of the results from \cite{Berenstein-Maldacena} were later checked in gauge theory calculations \cite{Drukker-Gross,Zarembo-Semenoff}, so we treat them as a strong coupling supersymmetric equivalent of (\ref{shif}). As a result of this calculation we discover the literal coincidence of the two ways of describing the gluon condensate in holographic models. 

Looking for the nonsupersymmetric version of the Wilson loop OPE we find that its form is rigidly determined by the dimension of the stack of the Dp-branes, which form the background of the model. Thus it has the same form in all models, based on D3-branes, being insensitive to the IR behavior of the model, and it is different in the Sakai-Sugimoto model \cite{Sakai,Witten_thermal}, surprisingly coinciding with the small coupling result (\ref{shif}). We note here, that the applicability of our approach to QCD is rather subtle, because the OPE is valid at a small enough size of the loop (namely $r \ll \la \alpha_s {\rm tr}(G^2) \ra^{-1/4}$), which implies a large enough energy scale. On the other hand, the supergravity description is valid only when the coupling constant is large, which in the case of QCD implies a small energy scale. Recall thought, that in QCD the coupling already becomes large at the scale of $1 {\rm GeV}$, which means that there is a window in the energy scale where our approach is applicable.  

In the second part of the paper we study the effect of the the gluon condensate on the Gross-Ooguri phase transition in the correlator of two circular Wilson loops. In a configuration with two small parallel loops of the same radius we evaluate the shift of the phase transition point due to the condensate. In a coplanar concentric configuration we rely on the numerical analysis and find an interesting change of the phase transition order, when radii of the loops become close to $\la \alpha_s {\rm tr}(G^2) \ra^{-1/4}$.

The work is organized as follows: in the first section we briefly review a simple holographic model that describes a nonzero gluon condensate and calculate the correction to the expectation value of the Wilson loop due to the presence of the condensate. In the second section we discuss the form of the OPE of a small Wilson loop in a general p-brane background and in the D4 Witten background \cite{Witten_thermal} in particular. Sections three and four are devoted to the study of the Gross-Ooguri transition in the presence of the condensate for parallel and concentric configurations, respectively. In the Appendix we provide the normalization procedure of the dilaton field in the model and the calculation of the gluon condensate, which is similar to \cite{decoupling}, but is included here for the sake of consistency and is rewritten in a slightly different notation.

\section{One loop, supersymmetric case}

We start by considering the holographic model based on the background of a stack of $N_c$ D3-branes. Its most famous feature is that in the throat limit this geometry looks like the $AdS_5 \times S_5$ space and exhibits a superconformal symmetry, the same as the $N$=4 SYM theory \cite{Maldacena_D3}. The supersymmetry can be partially broken by D7-branes, representing the fundamental quarks \cite{Karch_D3D7}, the conformal symmetry of $AdS_5$ can be violated by the introduction of various scale parameters \cite{Polchinski-Strassler, Liu-Tseytlin, Gubser, Constable}, but is restored in the vicinity of the boundary. As QCD is asymptotically free in the ultraviolet region (dual to the vicinity of the AdS boundary), we wish for the holographic dual of QCD to have this kind of boundary behavior. 
To study the gluon condensate in the model we need to take the supergravity solution with a nontrivial dilaton background (\ref{nontriv_dil}). A simple background, involving a nonzero gluon condensate is that of a D-instanton smeared on the D3-branes \cite{Liu-Tseytlin}. This background in the string frame has a metric
\begin{equation}
\label{D3_metric}
ds^2_{D3}|_{str} = \frac{L^2}{z^2} \sqrt{h_{-1}} (dx^\mu dx_\mu + dz^2 + z^2 d\Omega_5^2)
\end{equation}
and a dilaton
\begin{equation}
e^\phi =g_s h_{-1}, \qquad h_{-1} = 1 + \frac{q}{\lambda} z^4
\end{equation}
where $\lambda$ is related to the 't Hooft coupling constant of the dual field theory.
The curvature radius of the $AdS$ is related to the string coupling $g_s$ (the dilaton asymptotic value) as $L^4 = 4 \pi g_s N_c l_s^4$, where $l_s = \sqrt{\alpha'}$ is a string length scale. We see, that in the large $N_c$ limit this space is slightly curved and the supergravity description is valid. The dimensional parameter $q$ violates the conformal symmetry which is restored on the boundary $z \rar 0$.

The string coupling constant $g_s$ is related to the coupling constant of the dual Yang-Mills theory as follows \cite{Maldacena_D3, Ooguri-Oz}. 
\begin{equation}
g_{YM}^2 = 4 \pi g_s.
\end{equation}

To proceed with our study of the gluon condensate $\la {\rm tr}(G^2) \ra$ we need to fix the relation between the dimensional parameter $q$ and  $\la \alpha_s {\rm tr}(G^2) \ra$. This procedure has been performed in \cite{decoupling}, but we repeat it in the Appendix in a different notation, used in this paper, for the sake of consistency. The result is (\ref{trG}):
\begin{equation}
\label{trGphi}
\phi_4 = \frac{q}{\lambda} = \frac{ \pi^2}{\sqrt{2} \lambda } \frac{\la \alpha_s {\rm tr}(G^2) \ra}{N_c} 
\end{equation}

Now we can find the analogue of (\ref{shif}) in this model. In order to compute the Wilson loop via holography one needs to study the minimal area of the string worldsheet spanned on the Wilson contour at the boundary of the bulk space \cite{Maldacena_Wilson}
\begin{equation}
\label{wilson_recipe}
\la W(\mathcal{C}) \ra = e^{-Area(\mathcal{C})}.
\end{equation}
The area of the string worldsheet that we are interested in is described by the Nambu-Goto action in the string frame 
\begin{equation}
\label{NG}
 S_{NG} = \frac{1}{2 \pi l_s^2} \int d^2 \sigma \sqrt{g},
\end{equation}
where $g$ is the induced two-dimensional metric on the worldsheet.
In the string frame metric of the D3-model (\ref{D3_metric}) the action (\ref{NG}) for the circular Wilson contour of the radius $R$ takes the form (we expand the metric in $
\phi_4$, because $\phi_4 \sim \Lambda_{QCD}^4$ and we are considering the worldsheet in the region of the bulk space $z \sim R \ll \Lambda_{QCD}^{-1}$)
\begin{align}
\label{NGz}
S_{NG} &= \frac{2 \pi}{2 \pi l_s^2} \int \limits_0^R dr \ \frac{L^2}{z^2} \  r \sqrt{\dot{z}^2 + 1}  \ \left(1 + \frac{\phi_4 z^4}{2} + O(\phi_4^2 z^8) \right).
\end{align}
We assume here the parametrization of the worldsheet by the radius and angle variables and perform the integration over the latter due to the symmetry of the problem. At this stage we end up with a one-dimensional problem of finding a solution to the equation of motion, following from (\ref{NGz}). This solution at $\phi_4=0$ has a simple form  \cite{Drukker-Gross}
\begin{equation}
\label{z0}
z_0(r) = \sqrt{R^2 - r^2},
\end{equation}
and after the subtraction of a linear divergency gives a constant contribution to the area of the minimal surface.
We can calculate the correction due to the gluon condensate to the Wilson loop expectation value using the perturbation theory in $\phi_4 z^4$, and then plug this correction into the action (\ref{NGz}) in order to get in the linear order in $\phi_4 z^4$ (the superscripts denote the order in $\phi_4$) :
\begin{equation}
\label{variation}
\int \mathcal{L}^{0}(z_0 + z_1) + \mathcal{L}^{1}(z_0) = \int \left[ \mathcal{L}^{0}(z_0) + \mathcal{L}^{1}(z_0) + \left(\frac{\delta \mathcal{L}^{0}}{\delta z} - \p \frac{\delta \mathcal{L}^{0}}{\delta \p z} \right) z_1 \right] + \left. \frac{\delta \mathcal{L}^{0}}{\delta \p z} z_1 \right|_{boundary}
\end{equation}
The last term in the integral vanishes as it is the equation of motion, and the boundary term must vanish as well, because by definition the boundary value of the correction is 0. However, it does not in reality, because the expression $\frac{\delta \mathcal{L}^{0}(z_0)}{\delta \p z}$ diverges on the boundary when calculated on the zero-order solution, thus giving a finite input to the result. One can subtract this divergence by introducing some boundary counterterms, but we will use a somewhat more elegant approach. We shall simply rewrite everything in a different coordinate system, where this divergence do not exist at all. 

The form of the zero-order solution (\ref{z0}) gives us a hint on what coordinate system to choose. We will use the spherical coordinates, where the solution looks simply like a constant. We define them as
\begin{align}
z&=f \sin(\theta), \\
r&=f \cos(\theta), 
\end{align}
so that the zero-order solution is simply $f_0(\theta) = R$. The Nambu-Goto action (\ref{NGz}) in these coordinates is  
\begin{align}
\label{NGf}
S_{NG} &= \frac{L^2}{l_s^2} \int \limits_0^{\pi/2} d\theta \ \frac{\cos(\theta)}{f (\theta) \sin(\theta)^2} \   \sqrt{f' (\theta)^2 + f(\theta)^2}  \ \left(1 + \frac{\phi_4 }{2} f(\theta)^4 \sin(\theta)^4 \right).
\end{align}
One can check that the boundary term in (\ref{variation}) vanishes because $\frac{\delta \mathcal{L}^{0}(f_0)}{\delta \p f} = 0$. Thus we end up with a simple result coming from the full action, evaluated on the zero-order solution.
\begin{equation}
\label{one_loop}
S = \frac{L^2}{l_s^2} \int \limits_0^{\pi/2} d\theta \ \frac{\cos(\theta)}{\sin(\theta)^2} \     + \frac{\phi_4 R^4}{2} \cos(\theta) \sin(\theta)^2 .
\end{equation}
The first term, as expected, gives the usual linear divergency, proportional to the perimeter of the contour, while the second term defines the condensate part of the Wilson loop OPE:
\begin{equation}
\label{supershif}
\delta \la W \ra = - \frac{1}{6}  \frac{L^2  \phi_4}{l_s^2}  R^4 = - \frac{\pi^2 \sqrt{2} }{12 \sqrt{\lambda} } \frac{\la \alpha_s {\rm tr}(G^2) \ra}{N_c} R^4,
\end{equation}
where we have used the relation (\ref{trGphi}) and the definition of $L$. Note that we get the right sign of the correction and right $N_c$-dependence as compared with (\ref{shif}). Although the dependence on the coupling constant is not the same as in (\ref{shif}), this is not very surprising, because this result is computed in the strong coupling limit. If we compare our calculation of the Wilson loop in the presence of the gluon condensate with the computation of the OPE coefficients of a Wilson loop from \cite{Berenstein-Maldacena}, we find a literal coincidence. In \cite{Berenstein-Maldacena} the OPE is computed via the exchange of a dilaton between two spherical loops. It is presented as the integration of the dilaton vertex operator over two surfaces, multiplied by the propagator of the dilaton field. The vertex operator is nothing but the correction to the action due to the dilaton that we have denoted $\mathcal{L}^1$, and the propagator gives a factor $z^4$, which is present in the profile of the dilaton in our model. It turns out that by calculating the first order action on the zeroth order solution, we calculate the same integral as in \cite{Berenstein-Maldacena}. Thus what we get does coincide with the strong coupling version of the Wilson loop OPE, and we conclude that the two ways of defining the gluon condensate in holography are in fact identical.

\section{One loop, nonsupersymmetric case}
To describe QCD holographically one needs to introduce a gravitational background, where supersymmetry and conformal symmetry are broken. As we have already mentioned, in the models based on D3-brane background \cite{Gubser,Constable}, the OPE of the Wilson loop will look like (\ref{supershif}). But these models are not the only case. For instance, the Sakai-Sugimoto AdS/QCD model \cite{Sakai} is based on the D4-brane background, developed in \cite{Witten_thermal}. For the sake of generality, we will consider the case of $N_c$ p-branes here. For an arbitrary p this supergravity solution in the string frame looks like (\cite{p-branes1,p-branes2,p-branes3} see \cite{Ooguri-Oz} for a review)
\begin{equation}
ds_p^2|_{str} = h^{-\frac{1}{2}} \eta_{a b} dx^a dx^b +  h^{\frac{1}{2}} (du^2 + u^2 d\Omega^2_{8-p}),
\end{equation}
with a dilaton
\begin{equation}
e^{\Phi} = g_s h(u)^{-\frac{p-3}{4}},
\end{equation}
where in the throat limit
\begin{align}
h(u) &= \left( \frac{L}{u} \right)^{7-p}, \\
L^{7-p} &= d_p g_s N_c l_s^{7-p}, \\
d_p & = (4 \pi)^{\frac{5-p}{2}} \Gamma \left(\frac{7-p}{2} \right),
\end{align}
and indices $a,b$ denote the (p+1)-dimensional space parallel to p-brane. We can make the coordinate transformation \cite{Huang:2007fv}
\begin{equation}
\label{conf_coord}
u = \left( \frac{2}{5-p} \right)^{-\frac{2}{5-p}} L^{\frac{7-p}{5-p}} z^{-\frac{2}{5-p}}
\end{equation}
which brings the metric to a conformally flat form similar to (\ref{D3_metric})
\begin{equation}
ds_p^2|_{str} = \left( \frac{5-p}{2} \frac{L}{z} \right)^{\frac{7-p}{5-p}} \left( \eta_{a b} dx^a dx^b + dz^2 + \left(\frac{2}{p-5} \right)^2 z^2 d\Omega^2_{8-p} \right).
\end{equation}
Given this expression we can easily compute the induced metric on the string worldsheet with a circular boundary and hence the Nambu-Goto action (\ref{NG}).
\begin{equation}
S_{NG}|_{Dp} = \frac{1}{2 \pi l_s^2} \int \limits_0^R dr \  \left(\frac{5-p}{2} \frac{L}{z}\right) ^{\frac{7-p}{5-p}} \ r \sqrt{\dot{z}^2 + 1}
\end{equation}
We note, that the string coupling constant $g_s$ as well as $N_c$ enter this expression only via the parameter $L$. Making this dependence explicit, we find
\begin{equation}
\label{NGDp}
S_{NG}|_{Dp} \sim (g_s N_c)^\frac{1}{5-p} \sim \lambda^\frac{1}{5-p}
\end{equation}
We see now, that the dependence of the Nambu-Goto action on $\sqrt{\lambda}$ is rigidly related to the dimension of the underlying D-brane worldvolume. Let's consider the case of p=4. The string frame metric of this setting in the conformal coordinates (\ref{conf_coord}) is
\begin{equation}
\label{D4_metric}
ds^2_{str} = \left(\frac{L}{2 z}\right)^3 \left( \eta_{\mu \nu} dx^\mu dx^\nu +  f(z) d\tau^2 + \frac{1}{f(z)} dz^2 + 4 z^2 d\Omega^2_{4} \right).
\end{equation}
Here the $\tau$ dimension is compactified with the period $\delta \tau = \frac{16 \pi}{3} z_0$, the black hole warp factor is $f(z) = 1 - \frac{z^6}{z_0^6}$ and $\mu,\nu$ stand for usual 4D space indices. The dilaton is
\begin{equation}
e^\Phi = g_s  \left( \frac{L}{2 z}\right) ^{\frac{3}{2}}
\end{equation}
The parameter $L$ is related to the string coupling constant as
\begin{equation}
L^3 = \pi g_s N_c l_s^3  
\end{equation}
The relation between $g_s$ and the field theory coupling constant can be established similarly to the D3 case \cite{Kruczenski}:
\begin{equation}
\label{g_sD4}
g_s = \frac{2}{3 \pi} \frac{z_0}{l_s} g_{YM}^2. 
\end{equation}

Let us note here, that the metric (\ref{D4_metric}) is not conformal from the very beginning (see \cite{Nastase} for a review), as the rescaling of 4D coordinates can't be absorbed now in the rescaling of $z$. It is not asymptotically $AdS$ and its curvature behaves as $\frac{1}{z}$, making the classical gravity approach inapplicable near the boundary. Therefore the quantum gauge theory dual to this geometry is not asymptotically free. This is a significant obstacle for us, because now our method of the dilaton operator normalization (\ref{OO}) is useless. Indeed, in an asymptotically non-free theory the consideration of the leading divergence of a perturbation theory (\ref{GG}) is not legitimate even in the limit of asymptotically large momenta. Thus, if we consider a perturbation of the dilaton profile by a dimensional quantity $\phi_4$ (as in the preceding section), we cannot unambiguously fix its relation to the gluon condensate. 

Nevertheless, let us assume, that in QCD a simple relation between the dilaton profile and gluon condensate, similar to (\ref{trGphi}), holds at least parametrically. In this case, we can proceed in our calculations.

The Nambu-Goto action in the D4-brane background (\ref{D4_metric}) is (from now on we omit all numerical factors, as we are interested on parametrical dependence only), see (\ref{NGDp}),
\begin{equation}
\label{NGD4}
S_{NG}|_{D4} \sim (g_s N_c) l_s \int \limits_0^R \frac{r}{z^3} \sqrt{1 + f(z)\dot{z}^2}
\end{equation}
Note that the scale of compactification $z_0$, enters now the equation of motion for $z(r)$ via the warp factor $f(z)$. We do not know the solution to the classical equation of motion, but we can analyze it from the point of view of dimensional analysis. The integral in (\ref{NGD4}) has a dimension (-1) in the units of length. There are two scales, $R$ and $z_0$, that can contribute to this dimension, but in the limit $R \rar 0$ the integral must remain finite in order for it to agree with the leading term in (\ref{shif}). Therefore the value of the integral computed on the classical solution must have the form $\frac{1}{z_0} g(R)$, where $g(R)$ has no negative powers of $R$. If we introduce now the parameter $q/ \lambda$ of dimension (-4) in the dilaton profile, then, similarly to (\ref{supershif}), it will enter this function as $g(R) = const + \frac{q}{\lambda} R^4 + O(q^2 R^8)$. At the end of the day we can find the parametric structure of the gluon condensate correction to the small Wilson loop
\begin{equation}
\delta \la W \ra \sim (g_s N_c) \ \frac{l_s}{z_0} (\phi_4 R^4 + \dots) \sim \frac{\la \alpha_s {\rm tr}(G^2) \ra} {N_c} S^2,
\end{equation}
where we have used the expressions for $g_s$ (\ref{g_sD4}) and $\phi_4$ (\ref{trGphi}). Interestingly, it is exactly the behavior that has been presented in (\ref{shif}).

\section{Gross-Ooguri transition, parallel loops}  

When studying the impact of the dilaton on the Wilson loop vacuum expectation values it is quite natural to expand the consideration to Wilson loop correlators. Particularly it is very instructive to investigate the behavior of a correlator of two parallel Wilson loops of equal radii $R$ that are separated by a distance $l$ in a transversal direction, which we shall do in this section, and of concentric coplanar Wilson loops of different radii, which we are going to study in the next section.

As it has been proposed by Maldacena \cite{Maldacena_Wilson}, the Wilson loop correlator may be calculated in string theory and is related to the surface area of a string worldsheet stretched over two Wilson contours. Generally speaking, two types of the worldsheet solution are classically admissible -- a connected and a disconnected one, and we have to choose the one that has the minimal action (and thus the surface area) of the two. As it has been first discussed in \cite{Gross-Ooguri}, when distance between the contours increases and reaches a certain critical value $l=l_{c}$, a disconnected worldsheet solution becomes more preferable than the connected one:
$$ S_{conn.}(l_{c}) = S_{disc.} .$$

Thus the connected Wilson loop correlator undergoes a second order phase transition -- the Gross--Ooguri phase transition -- and vanishes. (Actually, the disconnected surfaces can still exchange propagating gravitational modes, the account of which transforms the phase transition into a crossover. However, these effects are suppressed by $N_c$ and will not be considered in our paper.) At an even greater value $l=l_*$ the connected solution becomes unstable and ceases to exist. An analogous transition is possible in the case of two concentric Wilson loops, when it is determined by the ratio of their radii. 

These effects in the pure $AdS_5\times S^5$ target space in the two possible configurations as described above have been analyzed in detail in \cite{Zarembo-G-O} and \cite{Olesen:2000ji} respectively. In the case of parallel loops that will be under our consideration in this section the critical value equals $l_{c} = 0.91 R$ and the connected solution becomes unstable at $l_* = 1.04 R$. The same transition has been studied in a nonconformal background numerically in \cite{Nian:2009mw}. In this section we shall determine how the particulars of the Gross--Ooguri transition in the case of parallel loops are modified by the presence of a gluon condensate. 


\subsection{Equations of motion and boundary conditions}

Let us consider a pair of Wilson loops in a Euclidean $R^4$ space, parametrized as 
\begin{align}
{\cal C}_1=&\{(x^1)^2+(x^2)^2=R^2, \ x^3=-l/2, \ x^0=0\}  \nonumber  \\
{\cal C}_2=&\{(x^1)^2+(x^2)^2=R^2, \ x^3=l/2, \ x^0=0\}. \nonumber 
\end{align}
Let us also denote $x^1=r\cos{\varphi},\ x^2=r\sin{\varphi},\ x^3=x,\ x^0=t$. 
The correlator of the Wilson loops is related to the Nambu-Goto action corresponding to the minimal surface stretched between the two Wilson loops in the $AdS_5\times S^5$ space with a dilaton \cite{Maldacena_Wilson}: 
\be\langle W({\cal C}_1)W({\cal C}_2)\rangle = \exp{(-S_{NG})}.\nonumber\ee 
The metric tensor of the $AdS_5$ part has the form \cite{Liu-Tseytlin}, (\ref{D3_metric}):
\begin{align}
ds^2 
=& \left(1+\phi_4 z^4\right)^{1/2} \frac{L^2}{z^2}(dz^2+dt^2+dx^2+dr^2+r^2d\varphi^2). \nonumber
\end{align}
Due to obvious symmetries, the minimal surface will be parametrized as
\be t(\tau,\sigma)=0,\ \ \varphi(\tau,\sigma)=\sigma,\ \ x(\tau,\sigma) = x(\tau),\ \ r(\tau,\sigma) = r(\tau),\ \ z(\tau,\sigma)=z(\tau).\nonumber \ee 
This defines the action
\be S_{NG} = \frac{L^2}{\alpha'}\int d\tau  \frac{r}{z^2}\lb z'^2+r'^2+x'^2 \rb^{1/2}\left(1+\phi_4 z^4\right)^{1/2} , \label{Action}\ee
where a prime denotes a derivative with respect to $\tau$. A minimal surface is one that satisfies the equations of motion for the action (\ref{Action}). The form of the action implies that the momentum $k$ along the $x$-coordinate is conserved.

We shall find the minimal surface solution perturbatively in $\phi_4$. 


\subsection{Zero-order solution in different coordinates}

We will first discuss the zero-order solution $(z_0(\tau), r_0(\tau), x_0(\tau))$, i.e. the solution without the dilaton. In the analysis below we will follow somewhat closely \cite{Zarembo-G-O}. Let us choose first $\tau=x$. The zero-order equations of motion yield:
\be (r_0^2+z_0^2+x_0^2)''_{xx} = 0\label{EoMSumRZ}. \ee
Let us recall the boundary conditions:
\ba 
z(-l/2)=z(l/2)=0,&\qquad& r(-l/2)=r(l/2)=0, \label{BCZR}\\
z'(0)=0,&\qquad& r'(0)=0.\label{BCZ'R'}
\ea
These boundary conditions are the same both for the full and for the zero-order solutions. The latter (\ref{BCZ'R'}) reflects the fact that the solution is symmetric with respect to the $x=0$ hyperplane. As a result, we obtain from (\ref{EoMSumRZ}, \ref{BCZR}, \ref{BCZ'R'}) that
\be
z_0^2 + r_0^2 + x_0^2 = R^2 + l^2/4 \equiv a^2 .\label{Sum2Const}
\ee
The coordinate $x$ is denoted here as $x_0$ for the sake of universality with respect to different coordinate systems.

Now we shall choose different coordinates, namely, $\theta=\tau$, and
\ba
z=f(\theta)\sin{\theta},\ \ r=f(\theta)\cos(\theta),\ \ x=x(\theta).\label{SphericalCoord}
\ea
Due to the symmetries of the solution, it is clear that the angular coordinate $\theta$ runs from $0$ to some value $\theta_0$ as $x$ varies from $-l/2$ to $0$, the region $0<x<l/2$ corresponds to
$\theta$ running back from $\theta_0$ to $0$. The boundary is located at $\theta=0$ and we shall determine $\theta_0$ later. In the new coordinates the action of a connected worldsheet may be rewritten as:
\ba
S_{conn.}=2\sqrt{\lambda}\int\limits_0^{\theta_0} d\theta\ \frac{\cos{\theta}}{f\sin^2{\theta}}\lb f^2 + f'^2 +x'^2 \rb^{1/2} \lb 1+\phi_4 f^4 \sin^4{\theta} \rb^{1/2},\label{SphAction}
\ea
where a prime denotes now a derivative with respect to $\theta$.

Zero-order equations of motion for the action (\ref{SphAction}) are:
\ba
&&\frac{\cos{\theta}}{f_0\sin^2{\theta}} x_0' \lb f_0^2 + f_0'^2 +x_0'^2 \rb^{-1/2} = k,\label{SphEomX}\\
&&\lb \frac{\cos{\theta}}{f_0\sin^2{\theta}} f_0' \lb f_0^2 + f_0'^2 +x_0'^2 \rb^{-1/2} \rb'\nonumber\\ 
&&= -\frac{-\cos{\theta}}{f_0^2\sin^2{\theta}} \lb f_0^2 + f_0'^2 +x_0'^2 \rb^{1/2} + \frac{\cos{\theta}}{\sin^2{\theta}} \lb f_0^2 + f_0'^2 +x_0'^2 \rb^{-1/2} .\label{SphEomF}
\ea
Eq. (\ref{SphEomX}) allows us to express $x_0$ through $f_0$:
\be
x_0'^2 = \frac{k^2 f_0^2 \sin^4{\theta}(f_0^2 + f_0'^2)}{\cos^2{\theta} - k^2 f_0^2 \sin^4{\theta}}.\label{SphRelX}
\ee
The boundary conditions are transformed into:
\ba
x(0) = -l/2,&& x(\theta_0) = 0,\label{SphBCX}\\ 
x'(\theta_0) &=& \infty, \label{SphBCX'}\\
f(0) = R,&& f'(\theta_0) = 0 .\label{SphBCF}
\ea 
The last boundary conditions in (\ref{SphBCX'}, \ref{SphBCF}) may be obtained from the analysis of the behavior of $d\theta/dx,\ df/dx$ in cylindrical coordinates. Also 
note that the zero-order relation (\ref{Sum2Const}) implies that $f_0^2+x_0^2=const=a^2$.  This means that we can replace the boundary conditions (\ref{SphBCF}) with the equivalent
conditions that are more suitable for numerical calculations:
\be
f(\tau_0) = a,\ f'(\theta_0) = 0 .\label{SphBCFNumerical}
\ee

Using Eq. (\ref{SphRelX}) and boundary condition (\ref{SphBCX'}) we obtain the expression for $\theta_0$ through $k$:
\be
\theta_0 = \arccos{\lb \frac{\sqrt{4k^2 a^2+1}-1}{2k a} \rb}\equiv \theta_0[ka].\label{SphDetTau0}
\ee
Substituting $x'$ from (\ref{SphRelX}) into (\ref{SphEomF}) we get a single nonlinear second-order differential equation for $f_0(\theta)$ with two boundary conditions (\ref{SphBCFNumerical}).
We shall solve it numerically, with the solution $f_0(\theta;l/R,ka)$ depending on two dimensionless parameters: known $l/R$ and for the moment unknown $ka$. If we substitute the numerical solution $f_0(\theta;l/R,ka)$ into Eq. (\ref{SphRelX}), we may now integrate it, and taking into account boundary conditions (\ref{SphBCX}), we obtain:
\be
l/2 = x_0(\theta_0) - x_0(0) =  
k\ \int\limits_0^{\theta_0[ka]} d\theta\ \sqrt{ \frac{f_0^2 \sin^4{\theta}(f_0^2 + f_0'^2)}{\cos^2{\theta} - k^2 f_0^2 \sin^4{\theta}} }[l/R,ka].\label{SphDetK}
\ee
This allows us to determine the yet unknown $ka$ for each value of $l/R$ and find the solution $f_0(\theta;l/R)$.


\subsection{The on-shell action}

We can define the zero- and first-order corrections to the action (\ref{SphAction}) as follows:
\ba
S^{0}_{conn.} &=& 2\sqrt{\lambda}\int\limits_0^{\theta_0} d\theta\ \frac{\cos{\theta}}{f_0\sin^2{\theta}}\lb f_0^2 + f_0'^2 +x_0'^2 \rb^{1/2},\label{SphZeroAction}\\
S^{1}_{conn.} &=& \sqrt{\lambda}\phi_4 \int\limits_0^{\theta_0} d\theta\ f_0^3 \cos{\theta}\sin^2{\theta} \lb f_0^2 + f_0'^2 +x_0'^2 \rb^{1/2} 
             + \ld\lbr \frac{\delta S^{0}_{conn.}}{\delta f'} f_1(\theta) + 
\frac{\delta S^{0}_{conn.}}{\delta x'} x_1(\theta) \rbr\rv_{\theta=0}^{\theta=\theta_0},
\label{SphFirstAction}
\ea
where ($f_0$,$x_0$) and ($f_1$,$x_1$) are the zero- and first-order solutions respectively.

An important property of the solution $f_0(\theta;l/R)$ is that its derivative $f'_0(\theta) \sim O(\theta^2)$ when $\theta \rightarrow 0$. This means that the boundary terms in
(\ref{SphFirstAction})

\be
\ld\frac{\delta S^{0}_{conn.}}{\delta f'}\rv_{\theta \rightarrow 0} = const,\qquad \ld\frac{\delta S^{0}_{conn.}}{\delta f'}\rv_{\theta \rightarrow \theta_0} = 0,\qquad
\frac{\delta S^{0}_{conn.}}{\delta x'} \equiv k.\label{SphSF'X'}
\ee
The first order solutions have to satisfy the following boundary conditions:
\be 
x_1(0) = x_1(\theta_0) = f_1(0) = 0.\label{SphBC1}
\ee
Combining (\ref{SphSF'X'}, \ref{SphBC1}) we conclude that the boundary terms in (\ref{SphFirstAction}) turn out to be zero. Hence, 
substituting $x'$ from (\ref{SphRelX}) into (\ref{SphFirstAction}) we get:
\ba
S^{0}_{conn.} &=& 2\sqrt{\lambda} \int\limits_0^{\theta_0} d\theta\ \frac{\cos{\theta}}{f_0\sin^2{\theta}} \lb \frac{f_0^2 + f_0'^2}{1-k^2 f_0^2 \frac{\sin^4{\theta}}{\cos^2{\theta}}} \rb^{1/2},\label{SphZeroS}\\
S^{1}_{conn.} &=& \sqrt{\lambda}\phi_4 \int\limits_0^{\theta_0} d\theta\ f_0^3 \cos{\theta}\sin^2{\theta} \lb \frac{f_0^2 + f_0'^2}{1-k^2 f_0^2 \frac{\sin^4{\theta}}{\cos^2{\theta}}} \rb^{1/2}.\label{SphFirstS}
\ea

Let us now make a comment on the choice of coordinates. In the papers \cite{Zarembo-G-O, Olesen:2000ji, Nian:2009mw} the calculations are carried out with $\tau=x$. In our case staying in those coordinates would lead to nonzero boundary terms in (\ref{SphFirstAction}) (see discussion in Section II). Moreover, those terms would contain a singularity proportional to the dilaton, which cannot be cancelled out by subtracting a term proportional to the contour perimeter. In order to make sense of the singular expressions in those coordinates we would have to add boundary counterterms in the spirit of holographic renormalization. The choice $\tau=\theta$ allows us to deal with finite and regular first-order corrections straightaway.


\subsection{The Gross--Ooguri phase transition: numerical results}

The Gross-Ooguri phase transition occurs when the Wilson contours are separated by such a great distance $l_{c}$ that the disconnected solution becomes preferable to the connected one:

\be
S_{conn.}(l_{c}) = S_{disc.},\label{DefGOTransition}
\ee
where $S_{disc.}$ is the action on the disconnected solution (\ref{one_loop}):

\be
S^{0}_{disc.} = 2\sqrt{\lambda} \int\limits_{0}^{\pi/2} d\theta\ \frac{\cos{\theta}}{\sin^2{\theta}}.\label{SphZeroSDisc}
\ee
The factor $2$ here stands for the two connected parts of the worldsheet surface. If we separate now the actions and the distances into zero- and first-order quantities, we obtain:
\ba
&& S^{0}_{conn.}(l_{c 0}) = S^{0}_{disc.},\label{DefGO0}\\
&& l_{c 1} = \frac{ S^{1}_{disc.} - S^{1}_{conn.}(l_{c 0}) }{ \partial S^{0}_{conn.}/\partial l (l_{c 0}) }. \label{DetL1}
\ea
Note that $l_{c 0}$ is determined from Eq. (\ref{DefGO0}) as a distance at which \textit{non-singular} quantity $S^{0}_{conn.}(l_{c}) - S^{0}_{disc.}$ equals zero. Indeed, using expressions in (\ref{SphZeroS}, \ref{SphZeroSDisc}) it may be rewritten as:
\ba
&& S^{0}_{conn.} - S^{0}_{disc.} = -2\sqrt{\lambda}\ \frac{1-\sin{\theta_0}}{\sin{\theta_0}} + \nonumber\\ 
&& 2\sqrt{\lambda} \int\limits_{0}^{\theta_0} d\theta\ \frac{1}{\sin^2{\theta}} \frac{ f_0'^2 \cos^2{\theta} + k^2 f_0^4 \sin^4{\theta} }
{ f_0 \lb \cos^2(\theta) - k^2 f_0^2 \sin^4{\theta} \rb^{1/2} \lb \lb f_0^2 + f_0'^2 \rb^{1/2} + 
f_0\lb 1-k^2 f_0^2\frac{\sin^4{\theta}}{\cos^2{\theta}} \rb^{1/2} \rb }.\label{SphZeroSReg}
\ea
We have already noted that $f'_0(\theta) \sim O(\theta^2),\ \theta \rightarrow 0$. Hence, the integrand in Eq. (\ref{SphZeroSReg}) is finite at $\theta=0$, and, although it is mildly divergent at $\theta=\theta_0$, the integral is finite. This implies that the derivative $\dfrac{\partial S^{0}_{conn.}}{\partial l} (l_{c 0}) = \dfrac{\partial  ( S^{0}_{conn.} - S^{0}_{disc.} )}{\partial l} (l_{c 0})$ is regular as well. Substituting $f_0(\theta;l/R)$ into the actions (\ref{SphZeroS}, \ref{SphZeroSReg}) for various ratios $l/R$ we can numerically determine that $S^{0}_{conn.}(l) - S^{0}_{disc.} = 0$ with a margin of error $10^{-6}$ when $l=l_{c 0} = 0.9047 \  R$ and the derivative in (\ref{DetL1}) equals:
\be
\frac{\partial S^{0}_{conn.}}{\partial l} (l_{c 0}) = (1.20 \pm 0.01) \sqrt{\lambda} R^{-1},\label{DetDerivNum}
\ee
as well as find the first-order correction to the action on a connected solution:
\be
S^{1}_{conn.}(l_{c 0}) = 0.26 
 \ \sqrt{\lambda} \phi_4 R^4.\label{DetS1Num}
\ee
Combining (\ref{one_loop}, \ref{DetL1}, \ref{DetDerivNum}, \ref{DetS1Num}) we obtain:
\be
l_{c 1} = (6.05\pm 0.06) \times 10^{-2} \phi_4 R^5 . \label{DetL1Num}
\ee
Thus we discover that the presence of the gluon condensate drives the point of the Gross-Ooguri transition to larger distances between the loops. An interesting question is how the presence of the dilaton affects the point of the onset of the instability $l_*$. However, it is impossible to determine it in a perturbative analysis. In the next section we shall investigate the concentric loop configuration numerically.


\section{Gross-Ooguri transition, concentric loops}  
To study the correlator of two concentric coplanar Wilson loops in the presence of the gluon condensate we cannot rely on the perturbation theory analysis, because in this case we cannot simply take a small enough radius of the loops to ensure the inequality $\phi_4 R^4 \ll 1$. Thus we will treat this case numerically. In this section we study the correlator of two concentric circular Wilson loops with radii $R_1$ and $R_2$, lying in the plane $x=0$. While we treat this problem only classically, other interesting stringy effects were studied in a similar setting in \cite{Gorsky-Zakharov} and the semiclassical approximation was inspected in a recent paper \cite{Tseytlin-Alday}. Generally speaking, there are two possible forms of the minimal surface, ending on this boundary: two disconnected half-spheres and one connected surface, similar to a half-torus. As it was pointed out in \cite{Olesen:2000ji}, the connected surface exists only when the distance between loop contours is not very large. In our numerical study we will discover that this is indeed the case.

\begin{figure}[ht!]
    \begin{center}
        \subfigure[]{%
            \label{torai}
            \includegraphics[width=0.5\textwidth]{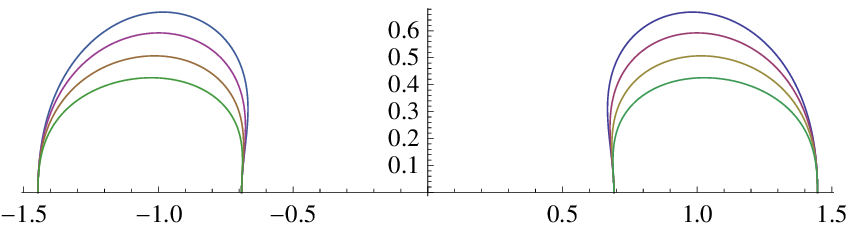}
        }%
        \subfigure[]{%
           \label{spheres}
           \includegraphics[width=0.34\textwidth]{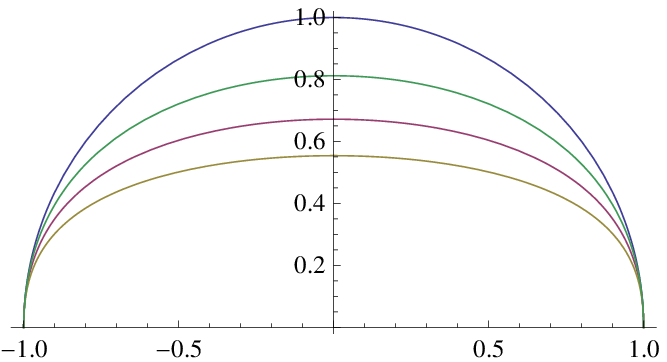}
        }%
    \end{center}
    \caption{%
        The minimal surfaces with boundary on two circles (a) and one circle (b) calculated at \hbox{ $\tilde{\phi}$ = 0, \ 5,\ 10, \ 40,} respectively from top to the bottom. 
     }%
\end{figure}

Considering the form of the boundary conditions (two concentric circles on the boundary of $AdS$) and the discussed above divergencies in the $(z,r)$-coordinates, it will be very convenient to adopt for our study the toroidal coordinates $(\tau,\sigma)$, defined as
\begin{align}
r(\tau,\sigma) &=\frac{ a \ \sinh(\tau)}{\cosh(\tau) - \cos(\sigma)} \\
z(\tau,\sigma) &=\frac{ a \ \sin(\sigma)}{\cosh(\tau) - \cos(\sigma)},
\end{align} 
where $a$ is a parameter of the coordinate system. The coordinate surfaces of constant $\tau=\tau_0$ are the tori with an axial circle of the radius $a \coth(\tau_0)$ and the radius of the section circle $\frac{a}{\sinh(\tau^0)}$. The coordinate $\sigma$ runs from $0$ to $2\pi$, being a kind of an angular variable in the section of the tori. We can choose the parameter $a$ in such a way that one of these tori satisfies our boundary conditions:
\begin{align}
\label{param}
a &= \sqrt{R_1 R_2}, \\
\cosh(\tau_0) &= \frac{R_1 + R_2}{R_1 - R_2}.
\end{align}
Hence our task is to find the minimal surface, described by the function $\tau(\sigma)$ with boundary conditions
\begin{equation}
\label{bound_tau}
\tau(0) = \tau(\pi) = \tau_0.
\end{equation}
The Nambu-Goto action (\ref{NG}) in toroidal coordinates looks like
\begin{equation}
\label{SNGtau}
S_{NG} = \int d \sigma  \ \sqrt{1 + \tau'(\sigma)} \ \frac{\sinh\left(\tau(\sigma)\right)}{\sin^2(\sigma)} \sqrt{1 + \phi_4 a^4  \left( \frac{  \sin(\sigma)}{\cosh(\tau) - \cos(\sigma)} \right)^4 }.
\end{equation}
Note that as we are not treating $\phi_4 a^4$ as a small parameter, we do not expand the square root factor in the metric (\ref{D3_metric}). We note here that before we have introduced the dimensional parameter $\phi_4$ the theory was conformal, and the only way the dimensional parameter $a$ enters the action is in the combination $\phi_4 a^4$. Thus we have 2 dimensionless parameters in the problem: $\tau_0$ (\ref{param}) and 
\begin{align}
\label{params}
\tilde{\phi} &= \phi_4 (R_1 R_2)^2.
\end{align}
We solve the equation of motion, following from (\ref{SNGtau}), with the boundary conditions (\ref{bound_tau}) numerically using the shooting method, obtaining the solutions at different $\tilde{\phi}$ (see Fig.\ref{torai}). To calculate the area of the corresponding surface, we regularize the action by subtracting the surface area of a torus with the same boundary: $\tau(\sigma) = \tau_0$. One can check, that the action on such a torus contains only the divergent term proportional to the perimeter. Hence by subtracting it we remove this divergency from the computed result.

At $\tilde{\phi} = 0$ we compare our results with the ones obtained in \cite{Olesen:2000ji} and find the agreement in the case when $R_1 = 2 \ R_2$, which is computed there. We also see that at a certain small $\tau_0 < \tau_{*}$, corresponding to large $R_1-R_2$ (\ref{params}), the connected solution does not exist. At $\tilde{\phi} = 0$ we find that $\tau_{*} = 1.405$, giving $R_1 = 2.72 \ R_2$. The Gross-Ooguri transition takes place when the regularized area of the connected surface becomes equal to $(-2)$ -- the area of the disconnected surface. In the case $\tilde{\phi} = 0$ we obtain the transition at $\tau_{c} = 1.5357$, or $R_1 = 2.40 \ R_2$.

To study the transition at $\tilde{\phi} \neq 0$, we need to solve a nonperturbative equation of motion for the disconnected surface as well. As it is just a pair of deformed half-spheres, we use the spherical coordinates like in (\ref{NGf}) and for different $\tilde{\phi}$ obtain numerical solutions, depicted on Fig.\ref{spheres}.

\begin{figure}[ht!]
    \begin{center}
        \subfigure[]{%
            \label{normalGO}
            \includegraphics[width=0.5\textwidth]{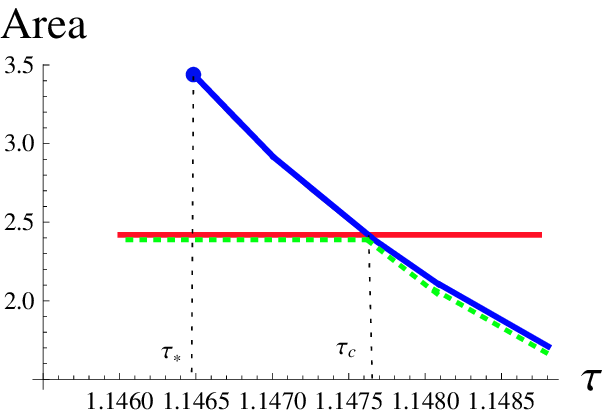}
        }%
        \subfigure[]{%
           \label{jumpGO}
           \includegraphics[width=0.4\textwidth]{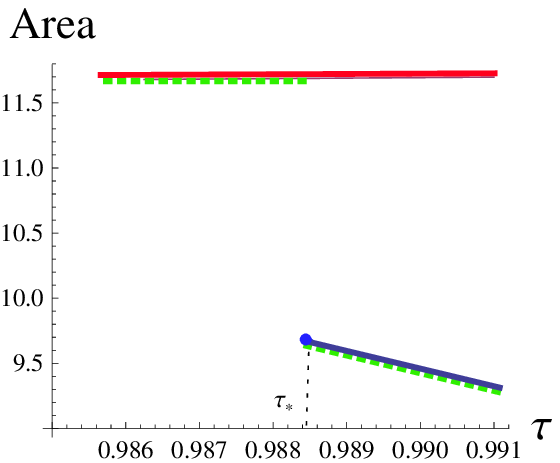}
        }%
    \end{center}
    \caption{%
        The area of the connected (blue) and disconnected (red) surface, depending on $\tau_0$: a)  at $\tilde{\phi} = 1$ the Gross-Ooguri transition is smooth, b)  at $\tilde{\phi} = 6$ the Gross-Ooguri transition has a jump. The green dashed line represents the minimal area.
     }%
\end{figure}

\begin{figure}[ht!]
    \begin{center}
        \subfigure[]{%
              \includegraphics[width=0.5\textwidth]{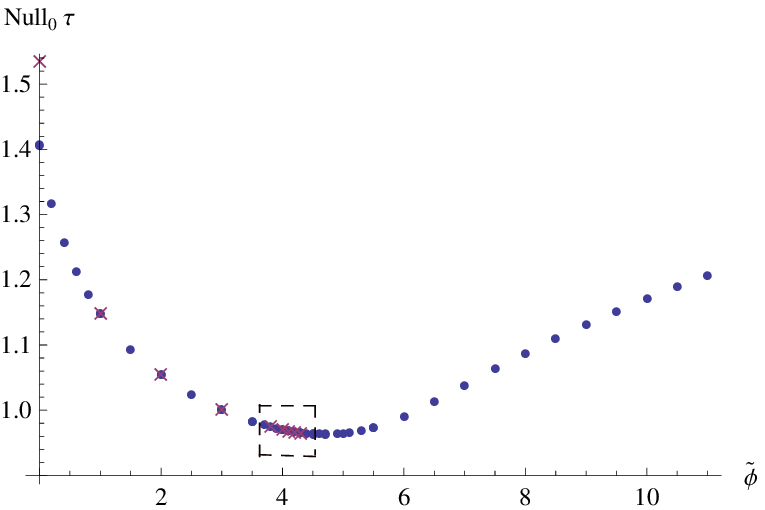}
        }%
        \subfigure[]{%
              \includegraphics[width=0.4\textwidth]{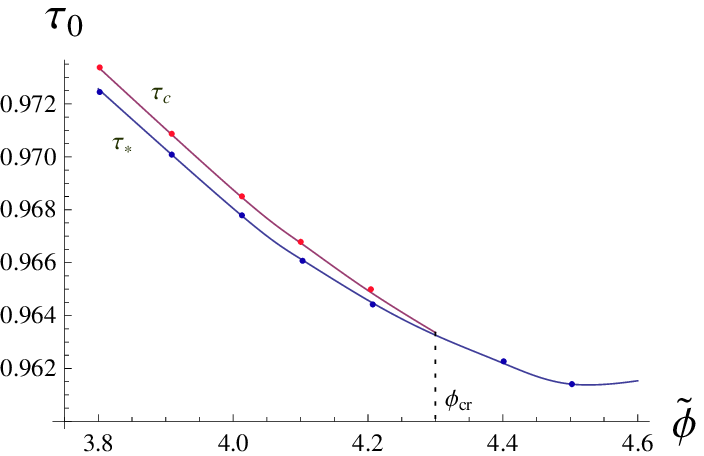}
        }%
    \end{center}
\label{phi_depend}
    \caption{%
        The dependence of $\tau_{*}$ (blue) and  $\tau_{c}$ (red) on $\tilde{\phi}$. a) the difference is not visible in this scale, b) enlarged dashed region of the first plot, the point, when the change of order occures is visible. 
     }%
\end{figure}

Depending on the value of $\tilde{\phi}$ we can look at two interesting points in $\tau_0$: the $\tau_{*}$ where the connected solution ceases to exist, and the $\tau_{c}$, where the area of the connected solution becomes larger than the area of the disconnected one. Obviously, $\tau_{c}$ cannot be less then $\tau_{*}$, because in that region there is no connected surface whose area can be measured. The case when $\tau_{c} > \tau_{*}$ is depicted in Fig. \ref{normalGO}. But as we saw in the preceding section, the account of the gluon condensate drives the point of the phase transition to a larger separation of the loops (\ref{DetL1Num}) and thus to smaller $\tau_{c}$, and at a large enough condensate it can reach the value $\tau_{*}$. Hence there can be a situation when the area of the connected surface remains less then the area of the disconnected all the way up to the point, where the connected surface ceases to exist, as it is depicted in Fig. \ref{jumpGO}. In this particular case the phase transition happens at $\tau_{*}$, and then the change of the form of the minimal surface is accompanied by a jump in the value of the minimal surface itself. Thus the Gross-Ooguri transition can change its order depending on the value of the gluon condensate. We find, that this is indeed the case and determine at what particular value of $\tilde{\phi}$ this change occurs. In Fig. \ref{phi_depend} we plot $\tau_{*}$ and $\tau_{c}$, determined at different values of  $\tilde{\phi}$. There is a point $\tilde{\phi}_{cr}$, where these curves intersect. It is located at $\tilde{\phi}_{cr} \approx 4.3$. Recalling our definition of dimensionless parameters (\ref{params}) we find that the characteristic size of the loops when the jump occurs is related to the gluon condensate as 
\begin{equation}
R \approx \la \alpha_s  {\rm tr} G^2 \ra^{-1/4}.
\end{equation}
Interestingly enough, we find that the coefficient is of order of
unity, so we can state that this change can indeed be relevant to the
usual hadronic physics at the energy scales less than 1 GeV.

\section{Conclusion}
In this paper we have studied the gluon condensate and its relation to the Wilson loops in holographic models. We have calculated the expectation value of a small circular Wilson loop in a supersymmetric holographic setting and found the agreement with the OPE of the Wilson loop, calculated in \cite{Berenstein-Maldacena}. Although in the nonsupersymmetric case of the holographic model, based on the compactified D4 Witten background \cite{Witten_thermal}, we have been unable to find exact results due to the nonconformality of the theory in the UV, nevertheless, we have obtained the parametric structure of the OPE coefficient under consideration. It turned out to be the same as in the weak coupling calculation \cite{Shifman}. 

Studying the effect of a nonzero condensate on the Gross-Ooguri phase transition in the correlator of two Wilson loops, we have discovered that it drives the point of the transition to a larger separation between the loops, which is quite natural, since the presence of the condensate makes the area of the disconnected surface larger. In the course of a numerical study of the correlator of two concentric loops we have found that at values of the condensate comparable to the order of the loop radii, the Gross-Ooguri transition changes its order, being accompanied by a jump not only in the derivative of the action, but in the area of the minimal surface itself. This is a new effect in the Wilson loop correlator behavior and it is worth further investigations. 

\acknowledgments
Authors are grateful to Alexander Gorsky for the supervision of their research.
A.K. would like to thank the Galileo Galilei Institute for Theoretical Physics for the hospitality and the
INFN for partial support during the completion of this work, as well as participants of the ``Large N gauge theories'' workshop for useful comments and discussions. A.K. is also grateful to Arkady Tseytlin and Konstantin Zarembo for insightful pieces of advice. The work of A.K. is partially supported by the Dynasty foundation, the grant RFBR-09-02-00308, and by the Ministry of Education and Science of the Russian Federation under contract 14.740.11.0347. The research of P.N.K. is partially supported by the Dynasty foundation, the grant RFBR-09-02-00308 and by the Ministry of Education and Science of the Russian Federation under contract 14.740.11.0081.

\appendix*
\section{Normalization of the dilaton}
Here we rewrite the procedure of the dilaton field normalization and the gluon condensate definition in the D3 model, used in \cite{decoupling}, in the notation of the present paper.  

Using the symmetry arguments we can state that the supergravitational dilaton field is dual to the gauge theory operator $O_\phi$, which is proportional to scalar gluon operator: $O_\phi = c_\phi  \alpha_s {\rm tr}(G^2)$. Hence, according to the standard AdS/CFT recipe \cite{Witten_AdS,Gubser-Klebanov} the boundary value of the dilaton field is treated as a source of $O_\phi$ and its normalizable mode as a vacuum expectation value $\la O_\phi \ra$. We can compute the two-point function $\la O_\phi O_\phi \ra$ to fix the coefficient $c_\phi$. In order to do this we need to calculate the classical action of the dilaton fluctuation and take the second variation with respect to its boundary value. The kinetic term of the dilaton field has a canonical form in the Einstein frame metric, which is defined as $G_{E} = \sqrt{g_s} e^{-\frac{1}{2}\phi} G_{S}$. In the Einstein frame the part of the supergravity action that we are interested in, is
\begin{align}
S_E &= \frac{1}{(2\pi)^7 l_s^8} \int d^{10}x \sqrt{G_E} \left(-\frac{1}{2} \p_A\phi \p^A \phi \right) \\
&= - \frac{N_c^2}{4 (2\pi)^2} \int d^4x dz \ \frac{1}{2 z^3} \left( (\p_z \phi)^2 + (\p_\mu \phi)^2 \right) 
\end{align}
In the last line we assumed, that there is no dynamics along the $S_5$-sphere, and used the definition of $L$ given above. The solution to the equation of motion near the boundary is
\begin{equation}
\phi(z,Q) = \phi_0 \frac{Q^2 z^2}{2} K_2(Qz), \qquad \phi(0,Q)=\phi_0,
\end{equation}
where $K_2$ is the McDonald function of the second kind. Substituting this solution back into the action and taking the second variation we get the expected two-point function. In the leading order of the large-$Q$ expansion it is exactly the conformal result
\begin{equation}
\label{OO}
\la O_\phi O_\phi \ra  =  \frac{N_c^2}{4 (2\pi)^2} \frac{1}{8} Q^4 \ln(Q^2 \epsilon),
\end{equation}
where $\epsilon$ is the $AdS$ space cutoff, which is related to the UV cutoff in the quantum field theory. This expression can be compared with the leading order in the large-$Q$ expansion of the correlator of scalar gluon operators, obtained in the QCD sum rule approach \cite{Kataev}
\begin{equation}
\label{GG}
\la {\rm tr}(G^2) {\rm tr}(G^2) \ra = \frac{N_c^2 - 1}{4 \pi^2} Q^4 \ln(Q^2 \epsilon^2),
\end{equation}
and this comparison in the large $N_c$ limit allows us to fix the operator, dual to the dilaton 
\begin{equation}
O_\phi = \frac{1}{4 \sqrt{2}} {\rm tr} G^2 .
\end{equation}
Similarly, we can compute the vacuum expectation value of $O_\phi$. As in the considered background the classical profile of the dilaton has a normalizable branch, $\phi = \frac{q}{\lambda} z^4$, we find a nonzero result
\begin{equation}
\la O_\phi \ra = \frac{N_c}{4 (2 \pi^2)} \phi(z,Q) \left. \frac{\partial_z\phi(z,Q)}{z^3} \right|_{z=\epsilon} = \frac{N_c}{\alpha_s (2 \pi^2)} q
\end{equation}
Thus, we find the desired expression for the gluon condensate in this model
\begin{equation}
\label{trG}
\la \alpha_s {\rm tr}(G^2) \ra \equiv 4 \sqrt{2} \alpha_s O_\phi = N_c \frac{4 \sqrt{2}}{(2 \pi)^2} q
\end{equation}
Note that the final relation doesn't include $\alpha_s$, and since $\la \alpha_s tr(G^2) \ra \sim N_c$, $q$ does not depend on $N_c$. Therefore all the dependencies on $N_c$ and $\alpha_s$ are reproduced by holography.

\end{document}